\documentclass[fleqn,10pt]{wlscirep}
\usepackage[utf8]{inputenc}
\usepackage[T1]{fontenc}
\title{On the origin of electron accumulation layer at clean InAs(111) surfaces}

\author[1,*]{Ivan Vrubel}
\author[1]{Dmitry Yudin}
\author[1]{Anastasiia Pervishko}
\affil[1]{Skolkovo Institute of Science and Technology, Moscow 121205, Russia}

\affil[*]{i.vrubel@skoltech.ru}

\begin{abstract}
In this paper, we provide a comprehensive theoretical analysis of the electronic structure of InAs(111) surfaces with a special attention paid to the energy region close to the fundamental bandgap. Starting from the bulk electronic structure of InAs as calculated using PBE functional with included Hubbard correction and spin-orbit coupling, we deliver proper values for the bandgap, split-off energy, as well as effective electron, light- and heavy-hole masses in full consistency with available experimental results. On the basis of optimized atomic surfaces we recover scanning tunneling microscopy images, which being supplied with accessible experimental data make it possible to speculate on the formation of electron accumulation layer for both As- and In-terminated InAs(111) surfaces. Moreover, these results are accompanied by band structure simulations of conduction band states.
\end{abstract}

\begin{document}

\flushbottom
\maketitle

\section*{Introduction}

Rising demands for increasing performance and continuous miniaturization of electronic devices have triggered a surge of technological interest to the morphology of III-V compound semiconductor surfaces. In particular, indium arsenide---a narrow bandgap semiconductor with exceptionally high electron mobility and carrier density---has been proposed for advancing various devices, including infrared lasers \cite{Miles1995,Zhang1995,Liu2004} and detectors \cite{Smith1987,Johnson1996,Mohseni2000}, photodiodes \cite{Fuchs1997,Rehm2005,Pour2011}, terahertz oscillators \cite{Rehm2013,Chang2013}, and single-molecule transistors \cite{Nacci2012,Martinez2015}. In the meantime, a further use of its unique electronic properties needs for a detailed understanding of surface structure of InAs. However, the study of even clean surfaces is a challenging task since they might exhibit a large variety of reconstructions; in other words, an ordering which minimizes the surface free energy, depending on the sample orientation as well as tiny details of manufacturing process, differs from the ideal bulk structure \cite{Hansson1988}. Clean InAs(111) surfaces are nowadays produced using diverse technological methods ranging from conventional molecular beam epitaxy  \cite{Andersson1994,Andersson1996surface,Taguchi2006}, ion sputtering \cite{Olsson1996}, atomic hydrogen \cite{Bell1998} and various wet treatments \cite{Ichikawa1998,Petrovykh2005,Tereshchenko2009}. It has been discovered that a stable In-terminated InAs surface exhibits a (2 $\times$ 2) In-vacancy structure \cite{Olsson1996,Yamaguchi1997,Bell1998,Taguchi2006}, whereas an As-terminated surface can appear in (1 $\times$ 1) unreconstructed and (2 $\times$ 2) reconstructed As ad-trimer configurations associated with the conditions of preparation \cite{Andersson1994, Andersson1996surface, Mankefors1999, Grishin2005, Taguchi2005, Tereshchenko2009}.

Typically, semiconductor materials host the active layer close to their surface explained by the pinning of the Fermi level at the bandgap region. In contrast, clean InAs surfaces \cite{Chen1989,Olsson1996,Mankefors1999} as well as upon impurity adatom adsorption \cite{Aristov1993,Aristov1994,Betti1998,Aristov1999,Morgenstern2000,Leandersson2003,Palmgren2005,Szamota2006,Szamota2007,Szamota2009,Szamota2011} exhibit natural electron accumulation in the near-surface area interpreted by the band bending and positioning of the Fermi surface above the conduction band minimum (CBM) \cite{Mead1963,Tsui1970,Laar1977,Chen1989, Smit1989,Noguchi1991,Aristov1993,Olsson1996Acc,Martinelli1997,Bell1998,Frost1999,Veal2001,Kanisawa2001,Mahboob2004,Piper2006}. This suggests that InAs surfaces provide a natural platform for the generation of quasi two-dimensional electron gas (2DEG) \cite{Morgenstern2002, Morgenstern2003}, where the electrons can move freely parallel to the surface while confined in other direction, and can serve as a favorable candidate for the development of novel low-dimensional quantum systems.

Tremendous progress in experimental methods supported by extensive theoretical studies has facilitated the practical use of InAs since the pioneering work of Tsui \cite{Tsui1970} . Nonetheless, the surface effects and the origin of accumulation layer formation are still debated. In this paper, we examine the stable atomic configurations of InAs(111) surfaces from first principles and propose the reasoning for generation of electron accumulation layer, which is supported by available experimental data based on angle-resolved photoemission measurements (ARPES) and scanning tunneling microscopy (STM).

\section*{Computational methods}

\begin{figure}[h!]
\centering
\includegraphics[scale=0.1]{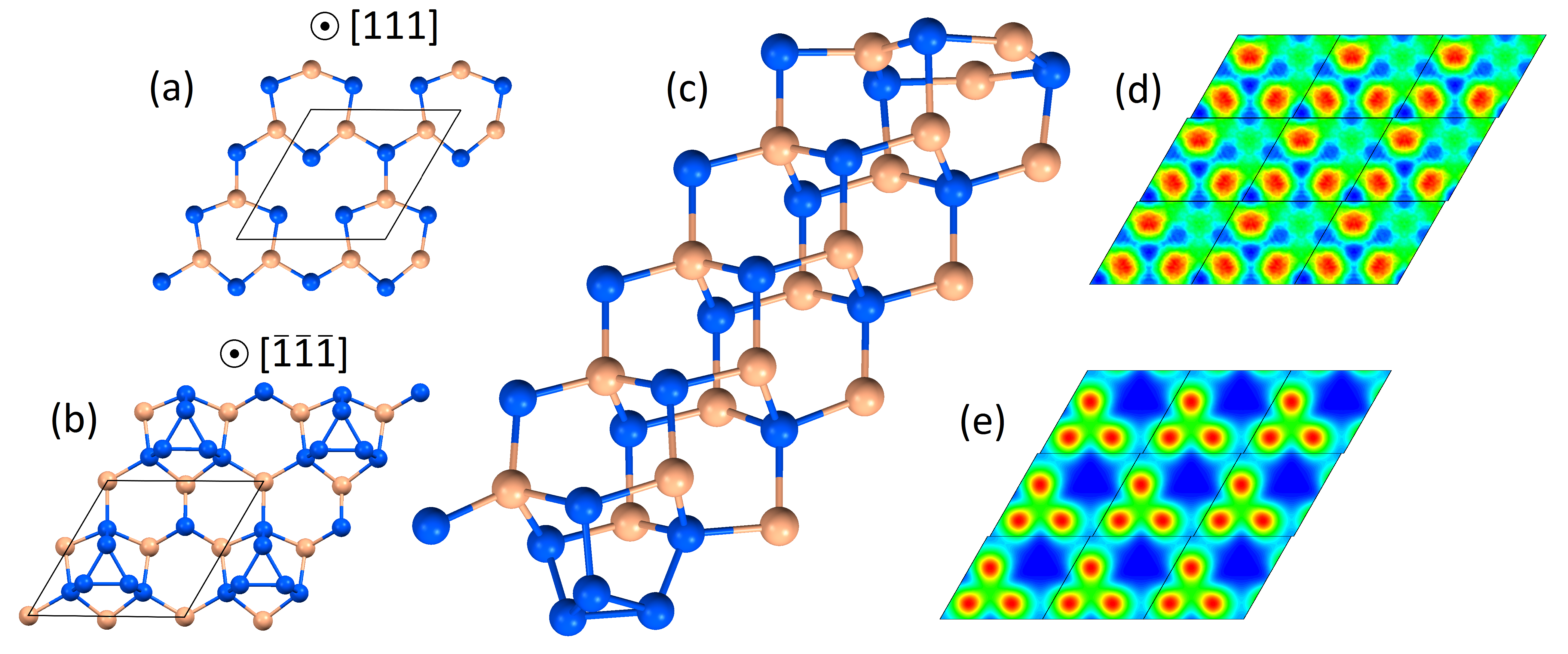}
\caption{(a) Top view of (2 $\times$ 2) reconstructed InAs(111)A surface, where the defect is formed by an In-vacancy. Indium atoms are depicted in yellow and arsenic atoms---in blue. Positions of the atoms in the supercell are optimized. The unit cell used throughout the numerical simulations is marked by black solid lines. (b) Top view of ad-trimer As structure for (2 $\times$ 2) reconstructed InAs(111)B surface. (c) Modelled slab geometry of InAs along the (111) direction.  (d) and (e) STM images as yielded by the numerical simulations of the (2 $\times$ 2) reconstructed InAs(111)A and InAs(111)B surfaces at the bias voltages of $V_\mathrm{bias}=-1$~V.}
\label{genview}
\end{figure}

We perform first-principles calculations using density functional theory (DFT) as implemented in Quantum ESPRESSO package \cite{qe09,qe17} based on the Perdew-Burke-Ernzerhof (PBE) generalized gradient approximation \cite{pbe96}. The plane-wave cutoff energy sufficient to reproduce material properties is set to 1000~eV. The InAs(111) structure is modelled by a supercell which reproduces a slab with 5 translations of the unit cell. The slab thickness along [111] direction is hence approximately 20~\AA, where the surfaces are given by As ad-trimer configuration from one side and In-vacancy reconstruction from the other, which is schematically shown in Fig.~\ref{genview}c. The slab is separated by a vacuum region of about 20~\AA\, thickness. Positions of the atoms are optimized and relaxed until the forces are less than 0.01~eV/\AA. Having been tested the total energy convergence we use a 3$\times$3$\times$3 automatic Monkhorst-Pack $k$-points set \cite{monkhorst76} to sample the Brillouin zone for self-consistent calculations and a 5$\times$5$\times$5 mesh for follow-up non-self-consistent runs.

A standard DFT approach to narrow bandgap semiconductors suffer from the bandgap problem and predict InAs to be a metal \cite{Cardona1987,Massidda1990}. This inadequate quantitative description of states within the bandgap region stimulated the use of more advanced approaches such as the hybrid-functional as well as $GW$ methods that have reached better results, however, demanding assortment of suitable functional or adjustment of self-energy operator for the correct description of InAs band structure \cite{Zhu1991,Lebegue2003,Chantis2006,Zanolli2007,kresse09}. An alternative approach is to introduce the additional on-site potentials applied to individual atomic sites (DFT+$U$ method) \cite{Zanolli2007,Weber2010, Vrubel2020}. Following recent studies, in this work we adopt this pseudohybrid Hubbard DFT methodology for InAs(111) slab calculations and find the optimal values of additional potentials by varying their strength until bulk InAs dispersion at the high-symmetry points closely meets experimental values \cite{Levinshtein1996,Vurgaftman2001}. For an accurate evaluation of the band structure we have also included the spin-orbit coupling (SOC) effect since zinc-blend type semiconductors are characterized by the degenerate valence band splitting in light-hole (LH), heavy-hole (HH), and split-off (SO) bands due to the lack of inversion symmetry \cite{Cardona1988}. Our numerical findings suggest that a reasonable accuracy can be achieved for the parameters $U_{\mathrm{In}}=0$ eV and $U_{\mathrm{As}}=3.3$ eV.

To make a direct comparison between numerical simulations and  available experimental data we evaluate the STM current using Tersoff-Hamann approach \cite{Tersoff1985}, where the variation of the tunneling current with the bias voltage $V_\mathrm{bias}$ in the system tip-substrate is proportional to the local density of states (LDOS) of the surface under consideration at the selected tip position, showing thus the impact from either occupied ($V_\mathrm{bias}<0$) or unoccupied ($V_\mathrm{bias}>0$) states in the energy window $ E_F - eV_\mathrm{bias}\leq E\leq E_F$. For a bias voltage probing occupied states the tunneling current can be expressed as:
\begin{equation}
I ({\bf R}) \propto \sum\limits_{\Psi_\mu:\;E_F-eV_\mathrm{bias}\leq E_\mu\leq E_F} |\Psi_\mu ({\bf R})|^2
\label{eq:I}
\end{equation}
here $I$ is the tunneling current, $\bf{R}$ stands for the position of the tip, $|\Psi_\mu ({\bf R})|^2$ is the local pseudocharge density, $V_\mathrm{bias}$ is the bias voltage between the tip and the substrate, while $E_F$ is the Fermi energy of the sample.

\section*{Results and discussion}
\subsection*{Electronic structure of bulk InAs}

\begin{figure}[h!]
\centering
\includegraphics[scale=0.3]{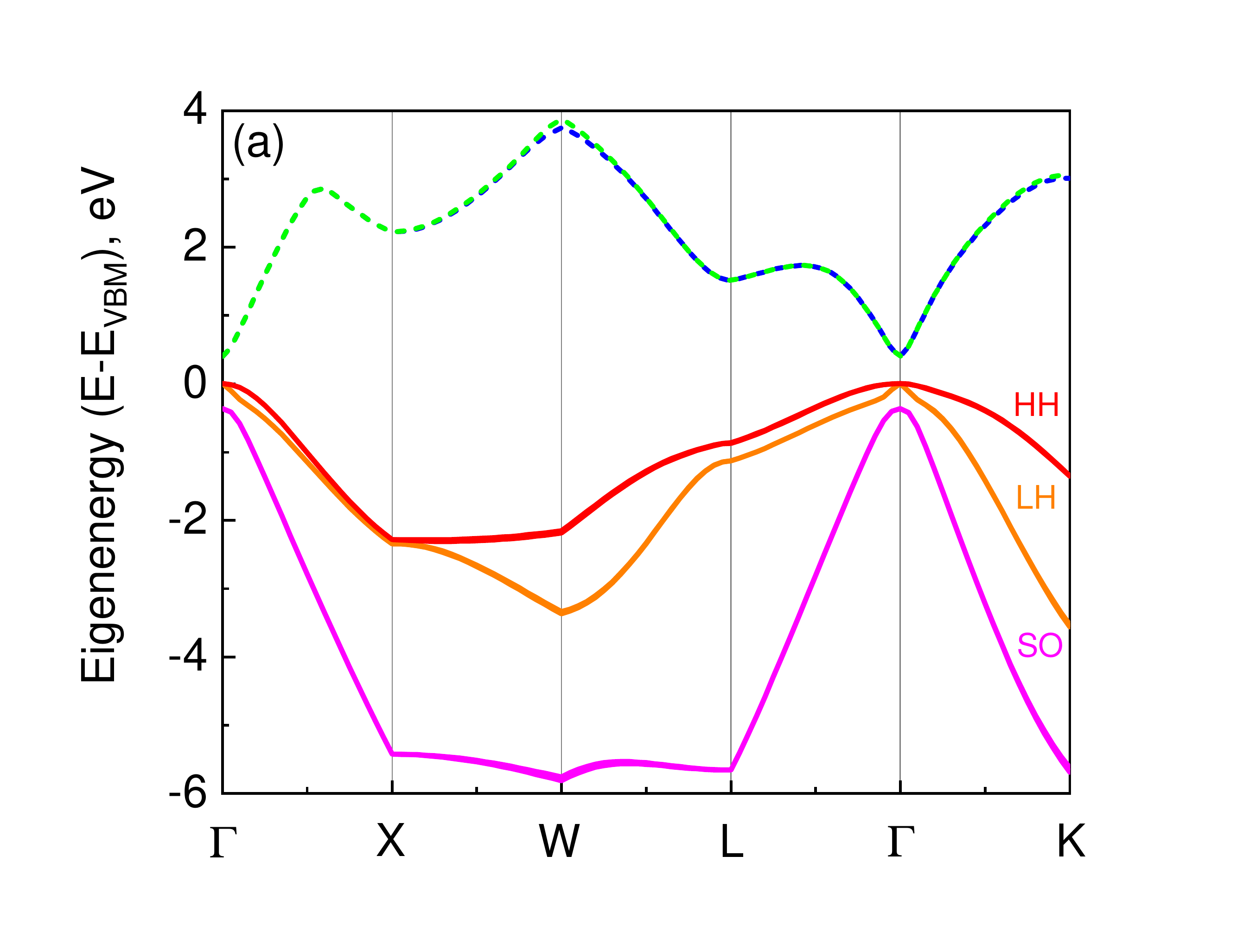}
\includegraphics[scale=0.3]{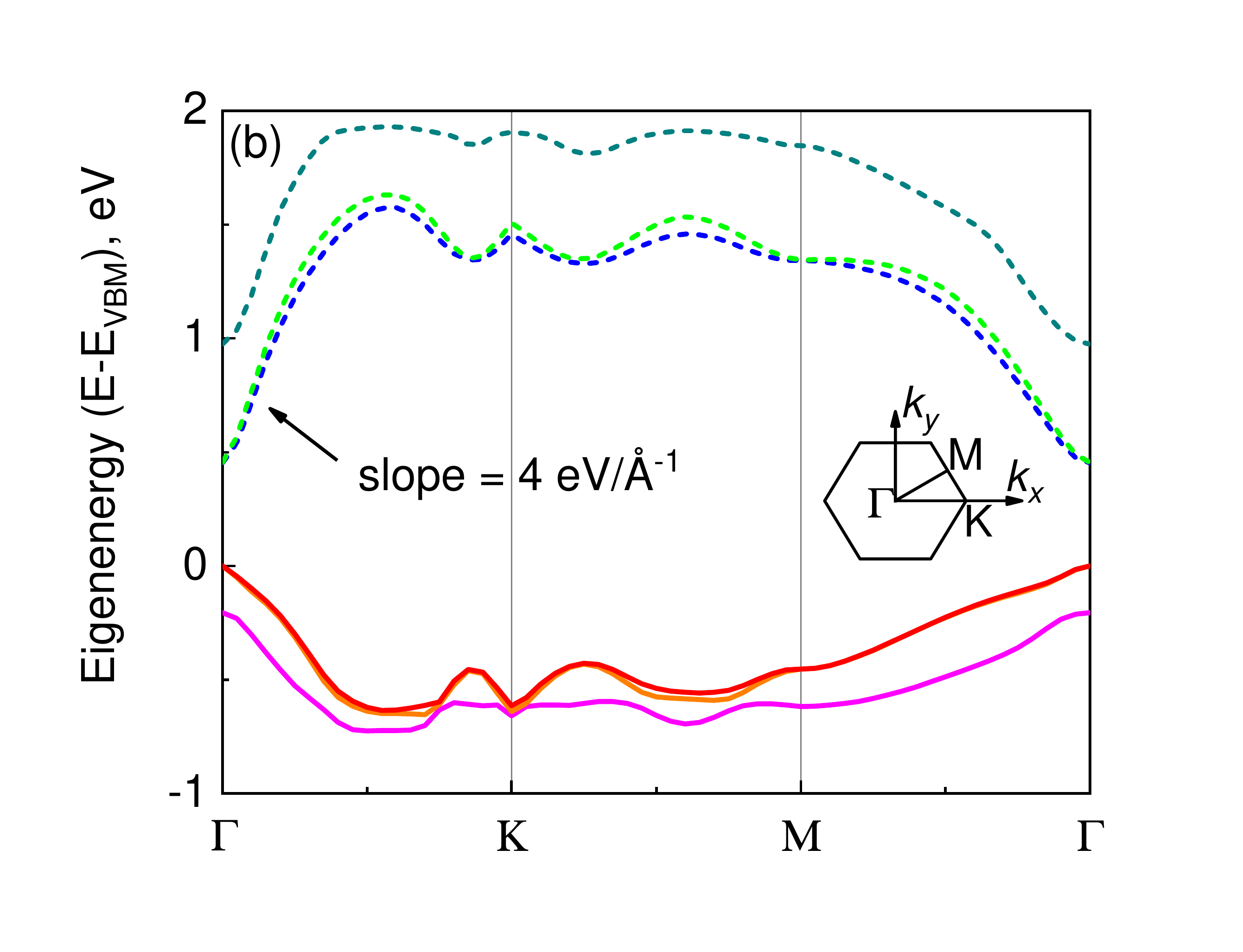}
\caption{(a) The band structure of InAs unit cell along $\Gamma$-X-W-L-$\Gamma$-K points as calculated using pseudohybrid Hubbard methodology with spin-orbit coupling. (b) The band structure of InAs supercell along $\Gamma$-K-M-$\Gamma$ path in the  reconstructed Brillouin zone, shown in inset.}
\label{band}
\end{figure}

First, we consider the bulk InAs illustrated in Fig.~\ref{genview}c. The optimized geometry of the system in the equilibrium is obtained with the lattice parameter 6.0584~\AA \; that matches the experimental value \cite{Vurgaftman2001}. Band structure for bulk material estimated with the use of pseudohybrid Hubbard method including SOC is shown in Fig.~\ref{band}a along high-symmetry points from the center of the Brillouin zone ($\Gamma$ point) with the coordinates $(0,0,0)$ to the points X (0,0.5,0), W (0.25,0.5,0), L (0.25,0.25,0.25), $\Gamma$, and K (0.375,0.375,0) in reciprocal space units. The valence band maximum (VBM) is set to zero energy. The $U$ potential value has been initially adjusted to reproduce the experimental bandgap value $E_g$, given in Table~\ref{Bulk}, and simultaneously resulted in the appearance of the appropriate spin-orbit splitting energy $E_{\mathrm{SO}}=0.37$ eV. When the spin-orbit effect is excluded in the calculations, the valence band is degenerate at the $\Gamma$-point similarly to conventional PBE results, whereas after inclusion of SOC, the states at the top of the valence band split into the distinctive light- and heavy-hole bands with corresponding effective masses, $m_{\mathrm{LH}}$ and $m_{\mathrm{HH}}$. Their values, extracted as the second order derivatives of the dispersion and given in Table~\ref{Bulk}, are also in good agreement with the available experimental data, however, the minor discrepancy of the obtained results with the experimental findings is explained by temperature dependence of band structure and its strong nonparabolicity in the vicinity of the $\Gamma$-point.

\begin{table}[]
\begin{tabular}{|p{6.0cm}|p{1.5cm} |p{1.5cm}|p{1.5cm}|p{1.5cm}|p{1.5cm}|}
\hline
Parameter & $E_g$ & $E_{\mathrm{SO}}$  & $m_{\mathrm{eff}}$  & $m_{\mathrm{HH}}$  & $m_{\mathrm{LH}}$ \\
\hline
Pseudohybrid Hubbard DFT with SOC &  0.40  & 0.37 &  0.032 & 0.320 & 0.039 \\
Experimental values  & 0.417 \cite{Lacroix1996} &  0.37 \cite{Zverev1979} & 0.026 \cite{kresse09} & 0.333 \cite{kresse09} & 0.027\cite{kresse09}  \\
\hline
\end{tabular}
\caption{Band parameters of bulk InAs at the $\Gamma$-point as compared to previously reported experimental data. All the energies are given in eV, whereas all the masses are in units of free-electron mass.}
\label{Bulk}
\end{table}

\subsection*{Clean InAs(111)A and InAs(111)B surfaces}

For purposes of clarity, in the following we assign indium-terminated surface to InAs(111)A and arsenic-terminated surface to InAs(111)B. During the slab geometry optimization, we deduce that the most stable surface configurations perform as $(2\times2)$ reconstructed surfaces, where InAs(111)A side is the surface with In-vacancy, shown in Fig.~\ref{genview}a, and InAs(111)B side is composed by As ad-trimers, Fig.~\ref{genview}b.

To validate structures with geometry specified in Figs.~\ref{genview}a,b and deduced within pseudohybrid Hubbard method we simulate the STM current using Tersoff-Hamann methodology as yielded by Eq.~\ref{eq:I} and make a direct comparison with available STM images. Results of STM measurements for InAs(111)A surface, discussed in Ref.\cite{Taguchi2006}, clearly manifest the intrinsic ($2\times2$) spatial surface irregularities on the distance of 4.2-4.4~\AA. The calculated STM map probing occupied states shows coherent results. The model of stable reconstructed surface is characterized by almost planar atomic configuration owing to an In-vacancy. The first atom of the reconstruction still has 4 nearest neighbors (natural for bulk), however is shifted from the tetrahedron closer to one of its faces. The rest 3 As atoms possess only 3 nearest neighbors. The bright three lobe pattern is related to the dangling bonds of these distorted ions. 

In contrast, the triangular shaped signals from As have been observed in experimental STM images for the opposite InAs(111)B surface, see e.g. Ref.\cite{Hilner2010}, which strongly supports As ad-trimer surface configuration. Theoretically predicted geometry of InAs(111)B with trimer morphology also appears in a simulated STM image, Fig.~\ref{genview}e, with the dominant contribution coming from As ad-trimers since they are located much higher than a pure As-terminated surface. Interestingly, it was recently argued in favor of the possibility of a stable $(1 \times 1)$ unreconstructed surface generation, depending on the preparation conditions \cite{Andersson1994,Hilner2010}, and even transition from $(2 \times 2)$ surface periodicity to $(1 \times 1)$ surface after annealing. Moreover, the presence of randomly distributed triangular shaped defects on the surface attributed to the missing As atoms was highlighted \cite{Grishin2005}. The stability analysis of this morphology suggests that InAs(111)B unreconstructed surface could be metastable \cite{Taguchi2005}. Upon modeling of $(1 \times 1)$ InAs(111)B structure, we observed that successful geometry optimization of a 4$\times$4$\times$4 mesh necessitates triangular shaped defects as yielded by As vacancies. Nevertheless, modelling of a supercell with randomly placed triangular defects is almost impossible due to its enormous size.

The $(2 \times 2)$ surface forms in-plane 2D hexagonal lattice with the doubled lattice constant as compared to the one for a bulk fcc InAs crystal, which leads to folding of the Brillouin zone. Notably, we can also neglect the dispersion relation in the direction normal to the InAs(111) surface due to the large thickness of a slab and find the dispersion along $\Gamma$-K-M-$\Gamma$ path within the reconstructed surface Brillouin zone shown in Fig.~\ref{band}b using pseudohybrid Hubbard density functional approach. It is worth mentioning that some features of the band structure demonstrated in Fig ~\ref{band}b merit closer attention. Interestingly, the electronic structure of a supercell still enjoys the bandgap of 0.5~eV at the $\Gamma$-point of the Brillouin zone. The curvature of the in-plane dispersion of unoccupied states was discussed in many experimental works \cite{Aristov1999, King2010, Olszowska2016}. Typically, one relies on ARPES technique which probes $E(k_\|)$ for occupied states directly. These states are populated due to thermally assisted charge separation resulting in the formation of the so called accumulation layer, which makes these initially empty states available for probing by ARPES. The experimental results unambiguously suggest that the band structure of occupied conduction band states in the vicinity of the $\Gamma$-point can be well fit with a parabola. The result of our numerical simulations is in agreement with the previously reported experimental value \cite{King2010, Olszowska2016}. It worth noting that a curvature of the conduction band in the vicinity of the $\Gamma$-point can be assessed \cite{King2008} using e.g. a non-parabolic approximation based on ${\bf k \cdot p}$ perturbation theory \cite{Kane1957}

\begin{figure}[h!]
\centering
\includegraphics[scale=0.3]{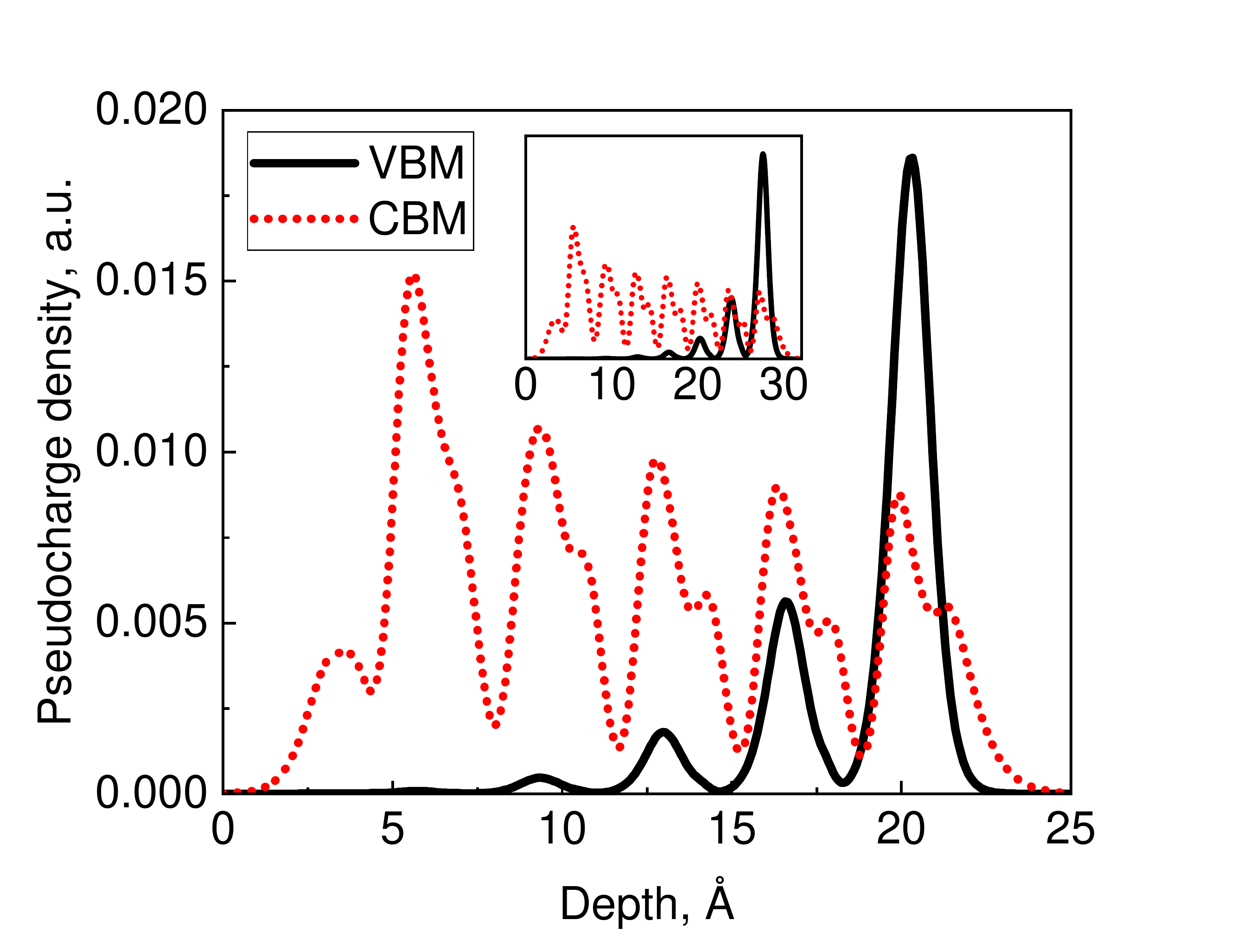}
\caption{Pseudocharge density formed by the states constituting the valence band maximum (VBM) and conduction band minimum (CBM) for InAs(111) slab of 5 unit cell layers. The pseudocharge density for the slab of 7 unit cell layers is provided in inset.}
\label{thin}
\end{figure}

Although there are several available studies  concerning InAs surface electronic structure \cite{Noguchi1991,Hakansson1997,Lowe2003,Aureli2005} the results are not consistent while discussing the surface band bending and the formation of charge accumulation states \cite{Tomaszewska2015}. Motivated by the study of accumulation layer formation on InAs(111) surfaces it seems reasonable to especially consider the distribution of pseudocharge densities given by the states constituting the VBM and CBM  at the $\Gamma$-point of reconstructed surface Brillouin zone, Fig.~\ref{thin}. From laser pump-probe photoemisson spectroscopy experiments the presence of charge accumulation layer in the near-interface InAs(111)A region has been attributed to the population of electron states above the VBM from the bulk states \cite{Grishin2005}. However, from our simulations one can notice that the states forming VBM are strongly confined in the vicinity of the InAs(111)A surface and subsequently can perform on the experiment as carrier accumulation layer at InAs(111)A side. In order to study the size effect on pseudocharge density distribution we also considered thicker slabs. In inset  Fig.~\ref{thin} we introduce the positions of molecular orbitals forming VBM and CBM of InAs(111) system composed by 7 layers of unit cells. As one can see, the molecular orbital redistributions for VBM and CBM do not depend on the slab thickness allowing InAs(111)A surface to accommodate charged carriers naturally in the $(2 \times 2)$ reconstructed surface configuration.

Previously reported ARPES experiments for InAs(111)B surface revealed the population of electron states in the conduction band leaded to the conclusion that electrons are strongly confined in reciprocal space resulting in the formation of an electron accumulation layer \cite{King2010}. Based on our theoretical findings, see e.g. Fig.~\ref{band}b, we can clearly distinguish a similar narrow conduction band that is unoccupied. The population of surface conduction states is explained by the downward band bending due to bulk and surface energy band adjustment, followed by positioning of the surface CBM below the Fermi level inherent to the bulk states. Previously, it was reported that VBM of surfaces states is positioned 0.56$\pm$0.05~eV below the Fermi level of bulk InAs\cite{Olsson1996} with the bandgap of 0.36~eV. For that reason, normally unoccupied conduction states become filled which result in the formation of 2DEG confined in the near-surface region. As it can clearly be seen in Fig.~\ref{thin} the orbitals forming CBM are positioned closely to InAs(111)B region and are able to confine thermally excited carriers coming from the bulk states due to the band bending effect in the near-surface region. However, the stable equilibrium, which can be observed by experimental methods, is dynamic and cannot be captured based on DFT approach. 

\section*{Conclusions}

In this paper, we studied the electronic properties of bulk InAs and clean InAs(111) surfaces using pseudohybrid Hubbard density functional approach with spin-orbit coupling. We evaluated the values for the bandgap and effective electron, heavy- and light-hole masses of the bulk material and showed them to be in good agreement with the existing experimental data. By modeling of STM images of the stable InAs(111) surface configurations we demonstrated that obtained surface reconstructions are in full consistency with the previously reported experimental results. In view of the experimental signs of accumulation layer formation on InAs(111) surfaces with $(2 \times 2)$ reconstruction we confirmed the appearance of 2DEG on InAs(111)B surface under population of surface conduction states. Whereas we attributed the experimental results on InAs(111)A side to redistribution of charged carriers in the system and their accumulation on that surface from the bulk states. It is worth noting that the calculated band structure of the CBM can be further utilized for modeling of the accumulation layer by means of coupled Poisson-Schr\"odinger equations.

\section*{Acknowledgements}
I.V. and D.Y. acknowledge the support from the Russian Foundation for Basic Research Project No. 20-52-S52001. The work of A.P. was supported by the Russian Science Foundation Project No. 20-72-00044 (analysis of the results of numerical simulations). The computations were enabled by resources provided by the ``Zhores'' supercomputer at Skolkovo Institute of Science and Technology.

\section*{Author contributions statement}

I.V. performed the numerical simulations. D.Y. and A.P. helped in the calculations and analyzed the obtained data. All authors discussed the results and contributed to the preparation of the manuscript.

\section*{Competing interests}
The authors declare no competing interests.

\bibliography{main.bbl}

\end{document}